%% file: iclr2026_conference.tex
\documentclass{article} 
\usepackage{iclr2026_conference,times}

\input{math_commands.tex}

\usepackage{hyperref}
\usepackage{url}
\usepackage{graphicx}
\usepackage{booktabs}
\usepackage{float}
\usepackage{diagbox}

\title{Learning Parameterized Nonlinear Elasticity on Curved Surfaces}

\author{Yankang Liu$^{1}$, Ke Zhang$^{2}$, Maziar Raissi$^{2}$, Roya Zandi$^{1}$
\\
$^{1}$ Department of Physics and Astronomy, University of California Riverside, USA\\
$^{2}$ Department of Mathematics, University of California Riverside, USA}

\iclrfinalcopy 
\begin{document}

\maketitle

\begin{abstract}
We learn parameterized nonlinear elasticity on curved surfaces using a physics-informed neural network that enforces governing equations and boundary conditions directly through the loss function, enabling a single trained model to represent a continuous family of elastic equilibria across geometric and material parameters. Nonlinear elasticity on curved manifolds underlies the mechanics of crystalline shells, elastic membranes, and viral capsids, where curvature and topological defects determine equilibrium structure and stability. Traditional exact and finite element solvers rely on symmetry reduction and must be reinitialized for each parameter choice, limiting scalability when symmetry is broken or parameters vary. We validate the proposed learning-based solver on a benchmark problem from curved elasticity, namely the one-dimensional single disclination on a spheroidal surface with known exact and numerical solutions. The network accurately reproduces these solutions, including parameter combinations excluded from training, demonstrating generalization across geometry and material regimes. This study establishes a scalable framework for learning nonlinear elastic systems on curved manifolds and lays the groundwork for extensions to fully two-dimensional and multi-defect configurations relevant to protein shells and other curved elastic networks.
\end{abstract}

\section{Introduction}
\begin{figure}[ht]
    \centering
    \includegraphics[width=0.8\linewidth]{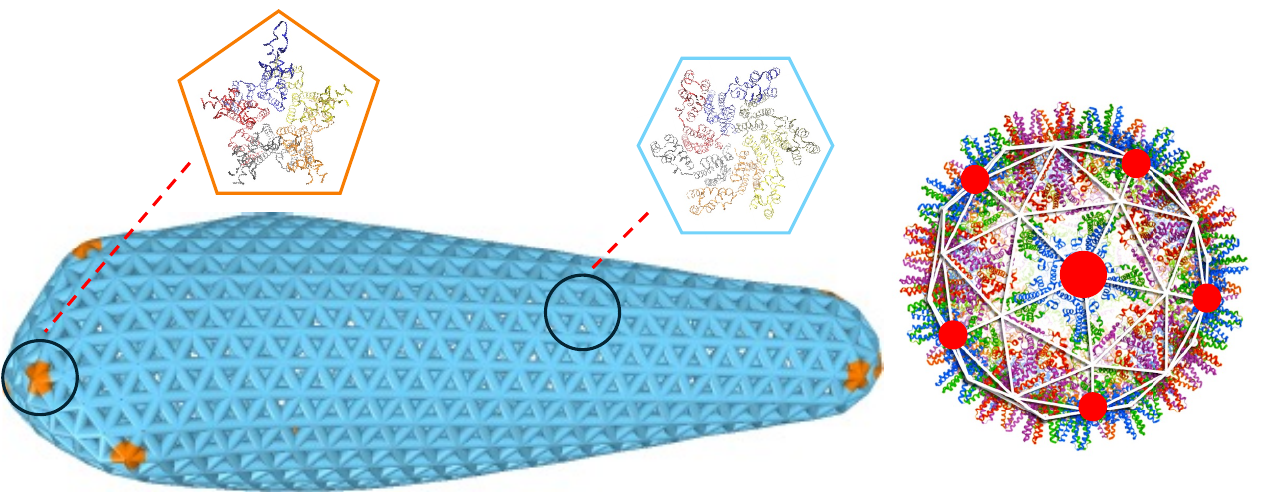}
    \caption{
Protein lattices on curved viral capsids. Icosahedral (HBV) and conical (HIV) capsids exhibit curvature-induced lattice defects that govern global shape and mechanical equilibrium. Red circles indicate pentameric defects required by curvature.
}
    \label{fig:virus}
\end{figure}
Nonlinear elasticity on curved surfaces governs the mechanics of crystalline protein shells, elastic membranes, and viral capsids \cite{Li2018,Li2019}. During viral assembly, proteins form curved lattices whose equilibrium structures can be understood as free energy minimizing elastic networks constrained by geometry. Curvature induces geometric frustration that requires topological defects, known as disclinations, in the lattice structure. The number and arrangement of these defects determine global shape and mechanical stability. Representative examples include icosahedral hepatitis B virus (HBV) capsids and the conical capsid of HIV, both of which exhibit ordered protein lattices on curved manifolds with defect-mediated geometry, see Fig.~\ref{fig:virus}.

Computing these equilibria requires solving nonlinear elasticity equations on curved manifolds. Exact and finite element methods are effective in highly symmetric configurations, such as a centered disclination on a spheroidal surface \cite{yinan2022}, where symmetry reduction yields a one-dimensional boundary value problem. However, when symmetry is broken, parameters vary, or multiple interacting defects are present, the system becomes fully two dimensional and classical solvers face scaling and stability limitations \cite{Liu2025}. Moreover, traditional approaches must typically be reinitialized for each new parameter choice, making exploration of continuous geometric and material regimes computationally expensive.

From a machine learning perspective, this setting corresponds to learning solutions of a geometry-dependent nonlinear partial differential equation with parameter-dependent metrics. While physics-informed neural networks (PINNs) have demonstrated promise across a range of differential equations \cite{Raissi2019}, most benchmarks focus on flat domains or relatively simple operators. It remains unclear whether physics-constrained learning can reliably handle nonlinear elasticity on curved manifolds and generalize across continuous parameter spaces.

In this work, we learn parameterized nonlinear elasticity on curved surfaces using a physics-informed neural network that enforces governing equations and boundary conditions directly through the loss function. The network is conditioned on geometric and material parameters, enabling a single trained model to represent a continuous family of elastic equilibria rather than solving isolated instances. We validate the approach on a benchmark problem consisting of a single centered disclination on a spheroidal surface with known exact and numerical solutions. Our goal is not to revisit the benchmark physics, but to establish a scalable learning-based solver that generalizes across parameter regimes and provides a foundation for extending physics-informed learning to fully two-dimensional and multi-defect configurations relevant to protein shells and other curved elastic networks.
\section{Problem Setup}
\subsection{Nonlinear Elasticity on a Curved Surface}
We consider nonlinear elasticity of a crystalline shell constrained to a curved surface. The deformation is described by a mapping between a reference configuration $\bar{x}$ and the actual surface $x$, inducing distinct metrics on the two manifolds. Mechanical equilibrium follows from minimizing the elastic energy under geometric compatibility, leading to the covariant force balance equation~\cite{Efrati2009, MosheSharonKupferman2015,Li2019}

\begin{equation} \label{eq:govern:formal}
\nabla_\alpha \sigma^{\alpha\beta}
+ (\bar{\Gamma}^\beta_{\gamma\nu} - \Gamma^\beta_{\gamma\nu})
\sigma^{\gamma\nu}
= 0,
\end{equation}

where $\sigma^{\alpha\beta}$ is the in-plane stress tensor and $\Gamma$, $\bar{\Gamma}$ denote the Christoffel symbols associated with the actual and reference geometries. This equation defines a nonlinear partial differential system whose coefficients depend explicitly on surface curvature and material parameters, resulting in a geometry-dependent nonlinear operator. We solve these equations subject to the boundary conditions $n_{\rho}\sigma^{\rho\lambda}\bar{g}_{\lambda \nu} + \tau n_{\nu}/r_A = 0$ and $\bar{\mathbf{r}}(\mathbf{r}=(0,0)) = (0, 0)$. See Appendix~\ref{Appendix:elasticity} for details.

In general the system is two dimensional. In this work, we focus on a symmetry-reduced benchmark consisting of a centered disclination on a spheroidal surface \cite{yinan2022}. Under rotational symmetry, the problem reduces to a one-dimensional boundary value equation for the radial deformation $\bar{r}(r)$, while retaining curvature-dependent nonlinear structure. Thus, although reduced in dimension, the governing equation remains geometrically nontrivial.

The equilibria depend on geometric and material parameters $(\beta, q, \nu_p)$, where $\beta$ controls spheroidal curvature (see Appendix~\ref{Appendix:spheroid}), $q$ denotes defect charge, and $\nu_p$ is the Poisson ratio. Varying these parameters generates a continuous family of elastic equilibria forming a parameterized solution manifold. Explicit forms of the reduced equations and boundary conditions are provided in Appendix~\ref{Appendix:elasticity}.

\subsection{Physics-Informed Learning of Parameterized Equilibria}

We approximate the radial deformation $\bar r(r)\big|_{(\beta,q,\nu_p)}$ using a deep fully connected neural network $u(r;\beta,q,\nu_p)$ with hyperbolic tangent activations. The spatial coordinate is $r \in [0, r_b]$, and $(\beta, q, \nu_p)$ denote geometric and material parameters. Automatic differentiation is used to compute derivatives with respect to the spatial coordinate and network parameters, while the parameters $(\beta, q, \nu_p)$ will not be differentiated. By explicitly conditioning on these parameters, a single network represents a continuous family of elastic equilibria rather than isolated parameter instances.


Let $F[\cdot]$ denote the nonlinear differential operator defined by the reduced equilibrium equation (\eqref{eq:govern:explicit}). The physical residual at a point $\tilde{r}\equiv(r, \beta, q, \nu_p)$ is denoted as $R(\tilde{r}) \equiv F[u(\tilde{r})]$ with the analogous definition at the boundary. Given $N$ sampled points, the loss functions are $\mathcal{L}_{\text{phys}}\equiv \sum_{i=1}^{N}|R(\tilde{r}_i)|^2/N$ and $\mathcal{L}_{\text{bc}}\equiv \sum_{i=1}^{N}|R(\tilde{r}_{bc,i})|^2/N$. The boundary condition at the origin is nothing but a regularization, which is enforced via $\mathcal{L}_{\text{0}}\equiv \sum_{i=1}^{N}|u(\tilde{r}_{0,i})|^2/N$. A small subset of ground truth data is incorporated through $\mathcal{L}_{\text{data}}\equiv \sum_{i=1}^N |u(\tilde{r}_i) - \bar{r}(r_i)|^2/N$. The total loss is the superposition of these contributions.

\section{Experiments}

We evaluate whether a single parameter-conditioned PINN can learn solutions across a continuous parameter space and remain stable under variations in collocation densities and network architectures.

Ground-truth solutions are generated using finite element method for $\beta\in[1,3]$, $q\in[0,1]$, $\nu_p\in[0.1,0.9]$ (step sizes are all 0.1), and 101 spatial points per configuration, yielding approximately $2\times10^5$ solution samples. Configurations with $\beta\in\{1,2,3\}$, $q\in\{0,1\}$, or $\nu_p\in\{0.3,0.8\}$ are excluded from training to evaluate interpolation across parameter space. Samples with $r<0.1$ are removed to avoid numerical instability near the origin. The size of the training set after the exclusion is around $0.57N$ where $N$ is the number of the sampled points from the whole data set.

\begin{figure}[ht]
  \centering
  \includegraphics[width=\columnwidth, clip]{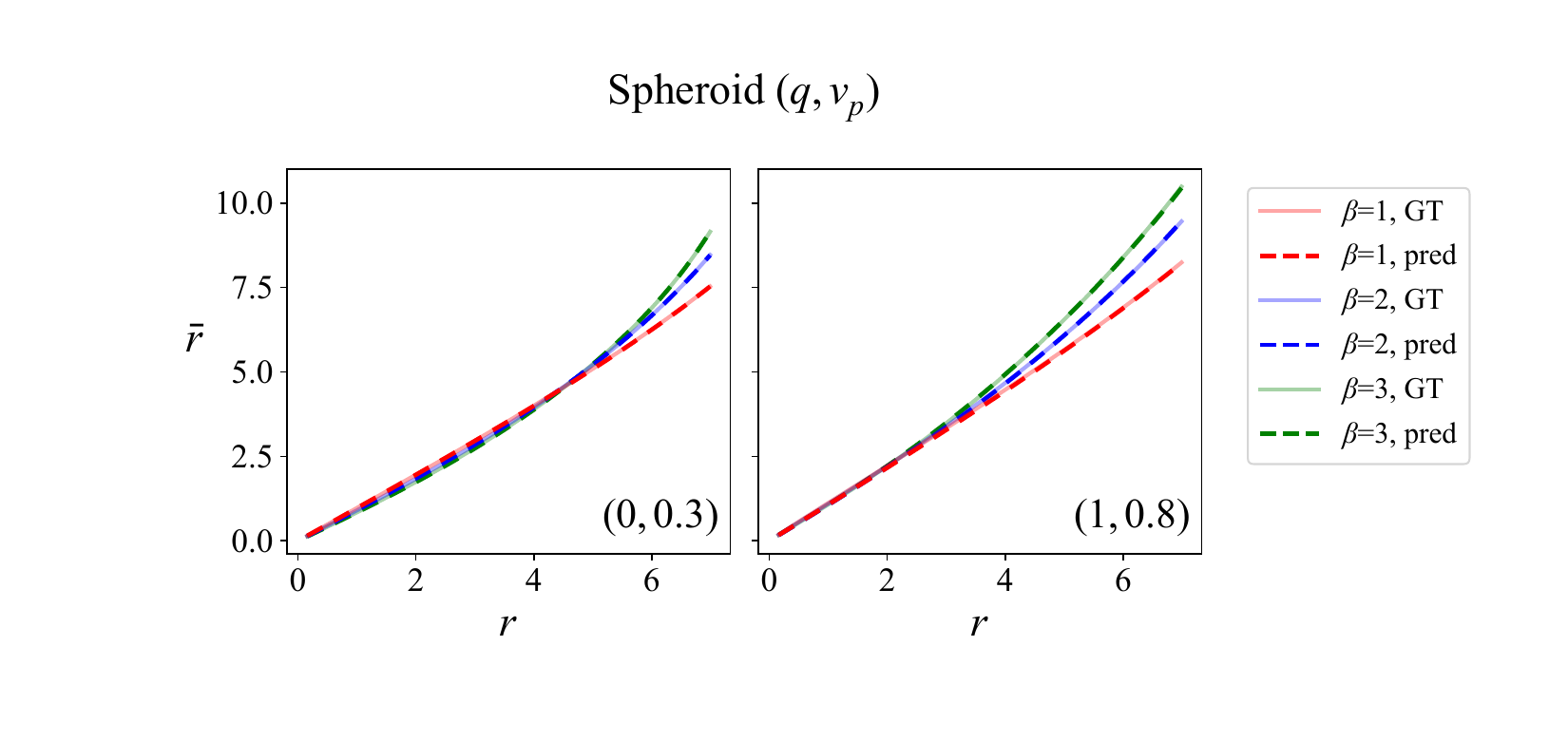}
  \vspace{-35pt}
  \caption{Comparison of neural network outputs (dashed lines) and ground truth solutions (light solid lines). Subplots show $(q,\nu_p) \in \{(0,0.3), (1,0.8)\}$. Colors red, blue, and green denote $\beta=1,2,3$, respectively. Setup: 4 layers, 200 neurons/layer, $N=4000$.}
  \label{lu}
\end{figure}

Figure~\ref{lu} compares neural network predictions with finite element solutions for parameter combinations excluded from training with $N=4000$. The model employs four hidden layers with 200 neurons per layers. Despite not being observed during optimization, the predicted deformation profiles exactly match the true solutions, indicating that the model captures the underlying parameterized solution structure rather than memorizing individual configurations. See Figure~\ref{fig:lb} for additional comparisons.

In the paper, all models are trained in a two-stage schedule: 200k optimization steps at learning rate $10^{-4}$ followed by 100k steps at $10^{-5}$. Training is implemented using PyTorch on two NVIDIA RTX A6000 GPUs. For the configuration shown above (Figure~\ref{lu}), the computational cost is approximately 6.2 seconds per 500 optimization steps. The early-stage convergence behavior is reported in Table~\ref{tab:phase1_log}.

We next examine sensitivity to the number of collocation points $N$ while keeping the same architecture. Table~\ref{tab:collocation} reports the total loss. Even with $N=1000$, corresponding to approximately 500 effective training samples after filtering, the loss remains on the order of $10^{-3}$, comparable to models trained with substantially larger $N$. This suggests that the physics-informed constraints significantly reduce the number of collocation points required to achieve high-precision solutions.

\begin{table}[t]
\centering
\caption{Total loss when varying collocation points under fixed architecture (4 layers, 200 neurons per layer). The loss value of the model in figure~\ref{lu} is highlighted in bold.}
\label{tab:collocation}
\begin{tabular}{lccccccc}
\toprule
$N$ & 1000 & 2000 & 3000 & 4000 & 5000 & 6000 & 7000 \\
\midrule
Loss & 1.619e-3 & 1.813e-3 & 1.765e-3 & \textbf{1.692e-3} & 1.652e-3 & 1.735e-3 & 1.677e-3 \\
\bottomrule
\end{tabular}
\end{table}

Finally, we investigate the effect of the network architecture by varying both the depth and width of the hidden layers while fixing the number of collocation points at $N=4000$. Table~\ref{tab:architecture} reports the corresponding loss values. While the majority of architectures converge to comparable error levels, severalconfigurations exhibit clear failure modes, characterized by large residual values ($\approx 22$), highlighting that appropriate architectural choices remain important for stable optimization.


\begin{table}[t]
\centering
\caption{Total loss when varying network architectures at $N=4000$. High values ($\approx 22$) indicate failure to converge. The loss value of the model in figure~\ref{lu} is highlighted in bold.}
\label{tab:architecture}
\begin{tabular}{cccccc}
\toprule
\diagbox{Neurons}{Layers} & 2 & 3 & 4 & 5 & 6 \\
\midrule
160 & 1.681e-3 & 2.251e0 & 2.251e0 & 2.252e0 & 1.679e-3 \\
200 & 1.700e-3 & 1.691e-3 & \textbf{1.692e-3} & 1.660e-3 & 1.665e-3 \\
240 & 1.690e-3 & 2.255e0 & 1.672e-3 & 1.670e-3 & 1.171e-3 \\
\bottomrule
\end{tabular}
\end{table}

Overall, a single parameter-conditioned PINN reproduces finite element solutions across a three-dimensional parameter space, and remains stable under moderate changes in collocation density and architecture.

\section{Conclusion}

We introduced a parameter-conditioned neural network for nonlinear elasticity on curved manifolds. By enforcing curvature-dependent equilibrium equations and conditioning on geometric and material parameters, the model learns a unified representation of a continuous family of elastic equilibria.

The results demonstrate accurate reconstruction of deformation profiles and stable interpolation across held-out parameter regimes within a single trained network. This parameterized learning formulation contrasts with conventional solvers, such as the finite element method, which require separate solutions for each parameter choice.

The symmetry-reduced benchmark retains essential geometric structure arising from curvature and defect-mediated mechanics, providing a principled testbed for geometry-aware physics-informed learning. These findings suggest that physics-informed neural networks (PINNs) offer a promising framework for scalable learning of geometry-dependent nonlinear systems on curved surfaces. The exploration on the full two-dimensional domain where multiple defects interacting with each other is a straightforward generalization of the current work and will be presented in forthcoming studies.

\subsubsection*{Acknowledgments}
Y.L. and R.Z. are supported by the National Science Foundation under Grant No.DMR-2131963 and MCB/PHY-2413062.

\bibliography{ref.bib}
\bibliographystyle{iclr2026_conference}

\appendix

\section{Elasticity Theory} \label{Appendix:elasticity}

This appendix presents the covariant elasticity theory and the explicit form of the governing \eqref{eq:govern:formal} and the boundary condition with $\tau=0$ under rotational symmetry.

The free energy of a partially formed elastic shell can be written as \cite{yinan2022}
\begin{equation}\label{eq:free_eng}
    \begin{aligned}
        F &= F^{\text{elastic}} + F^{\text{bending}} + F^{\text{abs}} + F^{\text{line}}\\
        &=\int d^2{\bm x}\sqrt{g}\left[{\cal F}^{\text{elastic}} + {\cal F}^{\text{bending}}\right]+F^{\text{abs}}+F^{\text{line}}\ ,
    \end{aligned}
\end{equation}
where the first and second terms are the stretching and bending energies, the third term represents the attractive monomer-monomer interaction promoting crystal growth, and the last term is associated with the cost of the line tension due to the presence of a boundary.
The elastic term ${\cal F}^{elastic}$ contains a quadratic term in the strain tensor $u_{\alpha\beta}$ (see \eqref{eq:strain:u}). The second term, in terms of the two radii of curvature $(R_i)_{i=1,2}$, is ${\cal F}^{bending} = \kappa[ (1/R_1-H_0)^2+ (1/R_2-H_0)^2]$ with $\kappa$ the bending rigidity and $H_0$ the mean spontaneous curvature of the subunits. The free energy density, \eqref{eq:free_eng}, has no trivial solution. The only surfaces allowing zero strains have either zero Gaussian curvature: a plane, a cylinder ($q=0$) or discrete delta function of Gaussian curvatures, like a cone ($q=1$). Surfaces with vanishing bending rigidity have a constant curvature radius $R_1=R_2=1/H_0$, forming a sphere. There is, therefore, no surface that simultaneously minimizes both the elastic and bending energies. The third term in \eqref{eq:free_eng}, $F^{\text{abs}}=-\Pi \hat{A} < 0$ with $\Pi$ the attractive interaction per unit area due to favorable hydrophobic contacts between subunits, is the driving force for crystal growth \cite{Morozov2010} and the last term is the cost of the line tension due to the presence of a boundary. 

The explicit form of ${\cal F}^{\text{elastic}}$ is
\begin{equation}\label{Eq:app:elastic energy density}
    {\cal F}^{elastic}=\frac{1}{2} A^{\alpha\beta\gamma\delta} u_{\alpha \beta} u_{\gamma \delta}\ ,
\end{equation}
where 
\begin{equation}
    u_{\alpha\beta} \equiv \frac{1}{2}\left[g_{\alpha\beta} - \bar{g}_{\alpha\beta}\right] \label{eq:strain:u}
\end{equation}
is the strain tensor and 
\begin{equation}\label{Eq:app:EF:Elastic_constant}
    A^{\alpha\beta\gamma\delta}=\frac{Y}{1-\nu^2_p}\left[\nu_p g^{\alpha\beta}g^{\gamma\delta}+(1-\nu_p)g^{\alpha\gamma}g^{\beta\delta} \right] \ ,
\end{equation}
with $Y$ the Young Modulus, $\nu_p$ the Poisson ratio and $g^{\alpha\beta}$ the inverse of the actual metric such that $g^{\mu\nu}g_{\nu\gamma}=g_{\nu\gamma}g^{\mu\nu}=\delta^{\mu}_{\gamma}$ where $\delta^{\mu}_{\gamma}$ is the Kronecker delta function. Since the Young’s modulus appears only as a prefactor, we choose $Y = 1$ throughout the paper. The stress tensor is
\begin{equation}\label{Eq:app:EF:Stress}
    \sigma^{\alpha\beta}\equiv\frac{1}{\sqrt{g}}\frac{\delta F}{\delta u_{\alpha \beta}}=A^{\alpha\beta\gamma\delta}  u_{\gamma \delta} \ .
\end{equation}

The Gaussian curvature is
\begin{equation}\label{Eq:app:EF:gaussian}
    K = \frac{\mbox{det}(\partial_i\partial_j f)}{(1+(\nabla f)^2)^2}=\frac{f^{\prime}(r)f^{\prime\prime}(r)}{r(1+f^{\prime}(r)^2)^2} \ ,
\end{equation}
and the mean curvature (with the convention that $R_i=R>0,i=1,2$ for the sphere) is
\begin{eqnarray}\label{Eq:app:EF:Mean}
    2H &=& -\nabla \cdot \left( \frac{\nabla f}{(1+(\nabla f)^2)^{1/2}} \right)\\ \nonumber
    &=&-\left( \frac{f^{\prime\prime}(r)}{(1+f^{\prime}(r)^2)^{3/2}}+\frac{f^{\prime}(r)}{r(1+f^{\prime}(r)^2)^{1/2}} \right)\ .
\end{eqnarray}
The two curvatures can be obtained from the equation
\begin{eqnarray}
    K&=&\frac{1}{R_1 R_2} \quad \nonumber\\
    2H&=&\frac{1}{R_1}+\frac{1}{R_2} \ ,
\end{eqnarray}
such that $1/R_1=H+\sqrt{H^2-K}$ and $1/R_2=H-\sqrt{H^2-K}$ with $H$ and $K$ given in Eqs.~\eqref{Eq:app:EF:gaussian} and ~\eqref{Eq:app:EF:Mean}.

Next, we provide various quantities for the surfaces of revolution, defined by $x = r\cos(\theta)$, $y=r\sin(\theta)$, $z = f(r)$ with actual metric
\begin{equation}
    ds^2 = \left[ 1 + f'(r)^2\right]dr^2 + r^2d\theta^2\ .
\end{equation}
The isotropic reference metric reads
\begin{equation}
    d\bar{s}^2 = \bar{r}'(r)^2dr^2 + \alpha^2 \bar{r}(r)^2d\theta^2 = d^2\bar{r}+\alpha^2\bar{r}^2d\theta^2\ .
\end{equation}
Note that due to the rotational symmetry, the angular mapping is identical, $\bar{\theta}(\theta)=\theta$

The nonzero Christoffel symbols are
\begin{equation}
\begin{array}{lccc}
\text{symbol}
& \Gamma_{rr}^{r}
& \Gamma_{\theta\theta}^{r}
& \Gamma_{\theta r}^{\theta} \\[6pt]
\text{reference}
& \dfrac{\bar{r}''(r)}{\bar{r}'(r)}
& -\alpha^2 \dfrac{\bar{r}(r)}{\bar{r}'(r)}
& \dfrac{\bar{r}'(r)}{\bar{r}(r)} \\[8pt]
\text{actual}
& \dfrac{f'(r)f''(r)}{1 + f'(r)^2}
& \dfrac{-r}{1 + f'(r)^2}
& \dfrac{1}{r}
\end{array}
\end{equation}
The elastic energy given in \eqref{eq:free_eng} depends on $\bar{r}(r)$,
\begin{equation}
\begin{aligned}
\frac{F^{\text {elastic }}}{Y \pi}= & \frac{1}{4\left(1-\nu_p^2\right)} \int d r r\left(1+f^{\prime}(r)^2\right)^{1 / 2} \\
& \times\left[\left(1-\frac{\bar{r}^{\prime}(r)^2}{1+f^{\prime}(r)^2}\right)^2+\left(1-\frac{\alpha^2 \bar{r}(r)^2}{r^2}\right)^2\right. \\
& \left.+2 \nu_p\left(1-\frac{\bar{r}^{\prime}(r)^2}{1+f^{\prime}(r)^2}\right)\left(1-\frac{\alpha^2 \bar{r}(r)^2}{r^2}\right)\right]
\end{aligned}
\end{equation}
The stress tensor given in \eqref{Eq:app:EF:Stress} becomes
\begin{equation}
\begin{aligned}
\sigma^{r r}= & \frac{Y}{2\left(1-\nu_p^2\right)\left(1+f^{\prime}(r)^2\right)}\left\{1-\frac{\bar{r}^{\prime}(r)^2}{1+f^{\prime}(r)^2}\right. \\
& \left.+\nu_p\left[1-\left(\frac{\alpha \bar{r}(r)}{r}\right)^2\right]\right\} \\
\sigma^{r \theta}= & \sigma^{\theta r}=0 \\
\sigma^{\theta \theta}= & \frac{Y}{2 r^2\left(1-\nu_p^2\right)}\left[\nu_p\left(1-\frac{\bar{r}^{\prime}(r)^2}{1+f^{\prime}(r)^2}\right)+1-\left(\frac{\alpha \bar{r}(r)}{r}\right)^2\right]\ .
\end{aligned}
\end{equation}
The general form of \eqref{eq:govern:formal} for $\beta=r$ becomes
\begin{equation}\label{eq:b4}
    \partial_r\sigma^{rr} + \bar{\Gamma}_{rr}^{r}\sigma^{rr} + \bar{\Gamma}_{\theta\theta}^{r}\sigma^{\theta\theta} + \Gamma_{rr}^r\sigma^{rr} + \Gamma_{\theta r}^{\theta}\sigma^{rr} = 0\ .
\end{equation}
The explicit form of the derivative of the stress tensor is
\begin{equation}
\begin{aligned}
\frac{d \sigma^{r r}}{d r}= & \frac{-Y}{\left(1-v_p^2\right)} \frac{f^{\prime}(r) f^{\prime \prime}(r)}{\left(1+f^{\prime}(r)^2\right)^2}\left\{1-\frac{\bar{r}^{\prime}(r)^2}{1+f^{\prime}(r)^2}\right. \\
& \left.+\nu_p\left[1-\left(\frac{\alpha \bar{r}(r)}{r}\right)^2\right]\right\} \\
& +\frac{Y}{\left(1-v_p^2\right)\left[1+f^{\prime}(r)^2\right]}\left[\frac{\bar{r}^{\prime}(r)^2 f^{\prime}(r) f^{\prime \prime}(r)}{\left[1+f^{\prime}(r)^2\right]^2}\right. \\
& \left.-\frac{\bar{r}^{\prime}(r) \bar{r}^{\prime \prime}(r)}{1+f^{\prime}(r)^2}+\nu_p \frac{\alpha^2 \bar{r}(r)^2}{r^3}-\nu_p \frac{\alpha^2 \bar{r}(r) \bar{r}^{\prime}(r)}{r^2}\right]\ .
\end{aligned}
\end{equation}
The equation determining $\bar{r}(r)$, \eqref{eq:b4}, becomes
\begin{equation}
\begin{aligned}
&-\frac{f'(r)f''(r)}{\left[1+f'(r)^2\right]^2} \left\{1-\frac{\bar{r}^{\prime}(r)^2}{1+f^{\prime}(r)^2}+\nu_p\left[1-\left(\frac{\alpha \bar{r}(r)}{r}\right)^2\right]\right\}\\
& +\frac{1}{1+f^{\prime}(r)^2} \times\left[\frac{\bar{r}^{\prime}(r)^2 f^{\prime}(r) f^{\prime \prime}(r)}{\left[1+f^{\prime}(r)^2\right]^2}-\frac{\bar{r}^{\prime}(r) \bar{r}^{\prime \prime}(r)}{1+f^{\prime}(r)^2}+\nu_p \frac{\alpha^2 \bar{r}(r)^2}{r^3}\right. \left.-\nu_p \frac{\alpha^2 \bar{r}(r) \bar{r}^{\prime}(r)}{r^2}\right]\\
&+\left(\frac{\bar{r}^{\prime \prime}(r)}{\bar{r}^{\prime}(r)}+\frac{f^{\prime}(r) f^{\prime \prime}(r)}{1+f^{\prime}(r)^2}+\frac{1}{r}\right) \times \frac{1}{2\left[1+f^{\prime}(r)^2\right]}\left\{1-\frac{\bar{r}^{\prime}(r)^2}{1+f^{\prime}(r)^2}+\nu_p\left[1-\left(\frac{\alpha \bar{r}(r)}{r}\right)^2\right]\right\} \\
& \quad-\frac{\alpha^2 \bar{r}(r)}{2 r^2 \bar{r}^{\prime}(r)} \times\left[\nu_p\left(1-\frac{\bar{r}^{\prime}(r)^2}{1+f^{\prime}(r)^2}\right)+1-\left(\frac{\alpha \bar{r}(r)}{r}\right)^2\right]=0\ .
\end{aligned} \label{eq:govern:explicit}
\end{equation}
The boundary conditions can be obtained through the variations of $F^{\text{area}} = F^{\text{elastic}} + F^{\text{bending}}$ in \eqref{eq:free_eng},
\begin{equation}
    n_{\rho}\sigma^{\rho\lambda}\bar{g}_{\lambda \nu} + \frac{\tau}{r_A}n_{\nu} = 0\ .
\end{equation}
For $\tau=0$, under rotational symmetry, the above equation becomes
\begin{equation}
    \begin{aligned}
        \frac{[\bar{r}'(r_b)]^2}{(1-\nu_p^2)\left\{(1+[f'(r_b)]^2\right\}}\left\{1 - \frac{[\bar{r}'(r_b)]^2}{1 + [f'(r_b)]^2} + \nu_p\left[1 - \left(\frac{\alpha\bar{r}(r_b)}{r_b}\right)^2\right]\right\}=0\ ,
    \end{aligned}\label{eq:bc1:explicit}
\end{equation}
where $r_b$ is the radius of the circular domain. We demand the correspondence between the center of the reference and the actual space, $\bar{\mathbf{r}}(\mathbf{r}=(0,0)) = (0, 0)$. In the reduced one-dimensional scenario, we have $\bar{r}(r=0)=0$.

\newpage
\section{Spheroid}\label{Appendix:spheroid}
\begin{figure}[ht]
  \centering
  \includegraphics[width=.7\columnwidth]{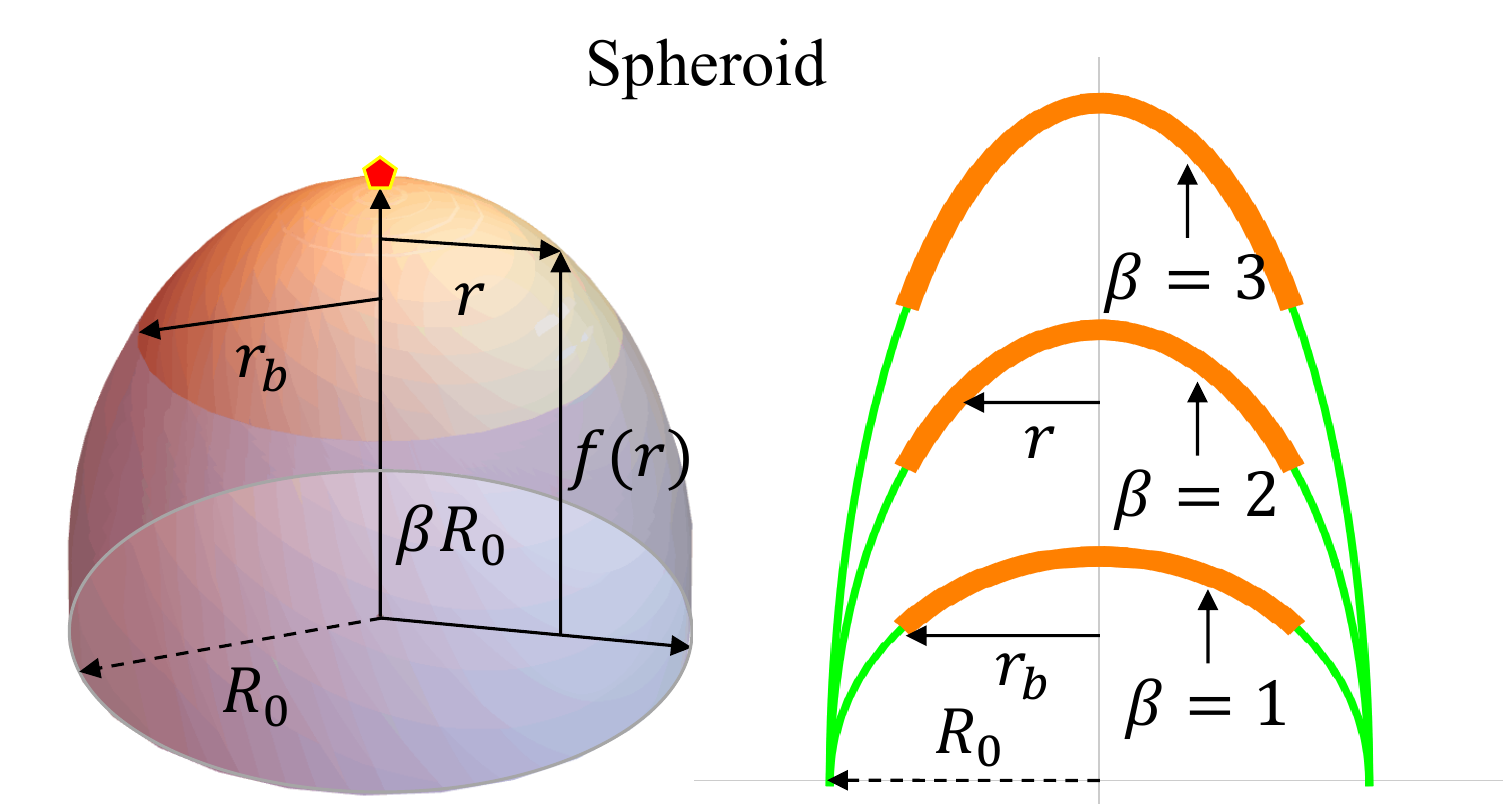}
  \caption{The figure presents the surface of revolution named spheroid, $f(r) = \beta \sqrt{R_0^2 - r^2}$, where $R_0$ is the radius of the spheroid, $r_b$ is the radius of the domain, and $\beta R_0$ is the height of the spheroid. As $\beta$ increases, the center of the spheroid goes higher and sharper. Note that $R_0$ has units of length and therefore sets the physical length scale. Throughout the paper, we fix $R_0 = 10$.}
  \label{fig:spheroid}
  \vspace{-10pt}
\end{figure}

\section{Additional Experiments}
\label{app:experiments}

\begin{figure}[ht]
  \centering
  \includegraphics[width=\columnwidth, trim={0 1cm 0 1cm}, clip]{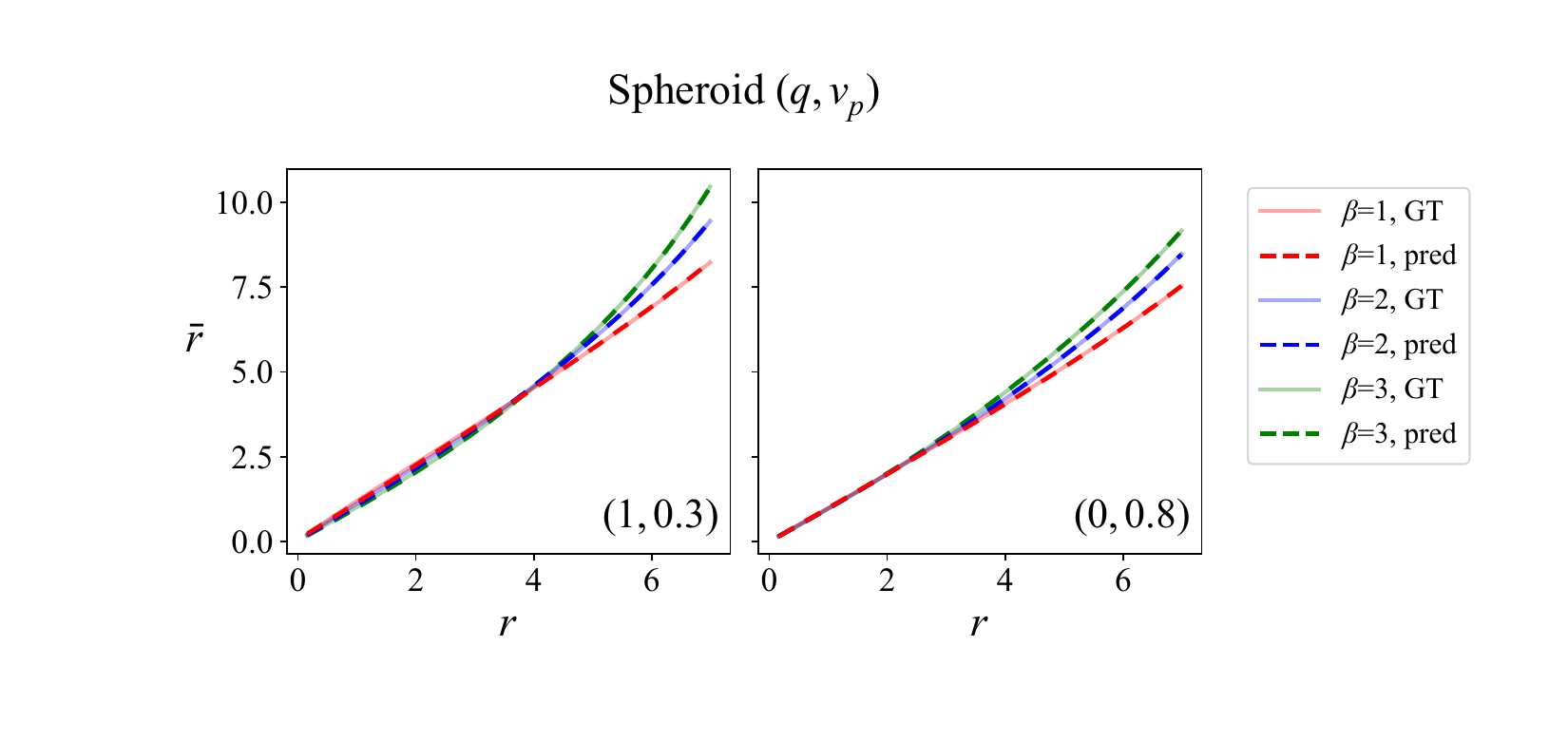}
  \vspace{-9pt}
  \caption{Comparison of neural network outputs (dashed lines) and ground truth solutions (light solid lines). Subplots show $(q,\nu_p) \in \{(1,0.3),(0,0.8)\}$. Colors red, blue, and green denote $\beta=1,2,3$, respectively. Setup: 4 layers, 200 neurons/layer and $N=4000$.}
  \label{fig:lb}
  \vspace{-10pt}
\end{figure}

We also include a small excerpt of the training log in Table~\ref{tab:phase1_log} for this configuration.
\begin{table}[ht]
\centering
\small
\setlength{\tabcolsep}{6pt}
\renewcommand{\arraystretch}{1.15}
\caption{Snapshot of stage 1 training log (every 500 epochs). Number of training points / $N$: 2167/4000. The neural network has 4 hidden layers and 200 neurons/layer. Time is per 500 epochs.}
\label{tab:phase1_log}
\begin{tabular}{r r r r r r}
\hline
Epoch & Loss & Data & Phys & BC & Time (s) \\
\hline
  500 & 21.981127 & 21.274858 & 0.706157 & 0.000111 & 6.53 \\
 1000 & 21.393473 & 21.134197 & 0.259183 & 0.000091 & 6.59 \\
 1500 & 21.124636 & 20.967300 & 0.157255 & 0.000080 & 6.34 \\
 2000 & 20.866369 & 20.746376 & 0.119921 & 0.000072 & 6.36 \\
 2500 & 20.547968 & 20.443615 & 0.104286 & 0.000066 & 6.20 \\
 3000 & 20.111891 & 20.014168 & 0.097665 & 0.000060 & 6.19 \\
 3500 & 19.467133 & 19.372320 & 0.094755 & 0.000057 & 6.21 \\
 4000 & 18.420986 & 18.327757 & 0.093162 & 0.000067 & 6.57 \\
 4500 & 16.492043 & 16.400759 & 0.091111 & 0.000173 & 6.22 \\
 5000 & 12.465515 & 12.379419 & 0.084866 & 0.001229 & 6.22 \\
 5500 &  5.312170 &  5.235003 & 0.066480 & 0.010686 & 6.21 \\
 6000 &  0.835386 &  0.757762 & 0.046624 & 0.031000 & 6.21 \\
 6500 &  0.467830 &  0.376380 & 0.071540 & 0.019910 & 6.20 \\
 7000 &  0.447981 &  0.352325 & 0.074996 & 0.020660 & 6.21 \\
 7500 &  0.430612 &  0.339458 & 0.069405 & 0.021749 & 6.21 \\
\hline
\end{tabular}
\end{table}

\end{document}

%% file: math_commands.tex
\usepackage{amsmath,amsfonts,bm}

\def\eqref#1{equation~\ref{#1}}

\def\1{\bm{1}}

\DeclareMathAlphabet{\mathsfit}{\encodingdefault}{\sfdefault}{m}{sl}
\SetMathAlphabet{\mathsfit}{bold}{\encodingdefault}{\sfdefault}{bx}{n}

